\documentclass[review]{elsarticle}

\usepackage{lineno,hyperref}
\usepackage{amsfonts}
\usepackage{amsmath,amssymb}
\usepackage{multirow}
\usepackage{xcolor}
\modulolinenumbers[5]

\journal{}

\begin{document}
 \def\lsim{\mathrel{\rlap{\lower4pt\hbox{\hskip1pt$\sim$}}
     \raise1pt\hbox{$<$}}}         
 \def\gsim{\mathrel{\rlap{\lower4pt\hbox{\hskip1pt$\sim$}}
     \raise1pt\hbox{$>$}}}         
\begin{frontmatter}

\linenumbers

\title{Solving  two-dimensional non-relativistic electronic and muonic atoms governed by Chern-Simon potential}


\author[mymainaddress,mysecondaryaddress]{Francisco Caruso}\corref{mycorrespondingauthor}
\cortext[mycorrespondingauthor]{Corresponding author}
\ead{francisco.caruso@gmail.com or caruso@cbpf.br}

\author[mymainaddress]{Jos\'{e} A.~Helay\"{e}l-Neto}
\author[mysecondaryaddress]{Vitor Oguri}
\author[mysecondaryaddress]{Felipe Silveira}

\address[mymainaddress]{Centro Brasileiro de Pesquisas F\'{\i}sicas -- Rua Dr.~Xavier Sigaud, 150, 22290-180, Urca, Rio de Janeiro, RJ, Brazil}
\address[mysecondaryaddress]{Instituto de F\'{\i}sica Armando Dias Tavares, Universidade do Estado do Rio de Janeiro -- Rua S\~ao Francisco Xavier, 524, 20550-900, Maracan\~a, Rio de Janeiro, RJ, Brazil}

\begin{abstract}
Bidimensional muonic and electronic atoms, with nuclei composed of a proton, deuteron, and triton, and governed by Chern-Simons potential, are numerically solved. Their eigenvalues and eigenfunctions are determined with a slightly modified Numerov method. Results are compared with those assuming that the same atoms are governed by the usual $1/r$ potential even in a two-dimensional space, as well as with its three-dimensional analogs.
\end{abstract}

\begin{keyword}
muon physics\sep muonic molecule \sep planar physics \sep quantum physics \sep \sep Numerov method.
\end{keyword}

\end{frontmatter}

\section{Introduction} \label{intro}
Currently, the description of a system similar to a hydrogen atom, a positive nucleus with an orbital negative particle, is one of the best developed both theoretically and experimentally. Since 1949 \cite{wheeler}, the interest in studying atoms in which the negative particle is no longer an electron but a muon -- the so-called muonic atom -- is growing up. These atoms, composed of only two bodies, are an excellent starting point for the theoretical study of new types of interactions \cite{Atabek, Eveker, Reiser, felipe, felipe2, felipe3}.

The interest in studying muonic hydrogen atom got a remarkable boost in 2010, after Pohl \textit{et al}.~\cite{Pohl,Antognini} reported on their outstanding result for the proton radius by the measurement of the Lamb shift in this category of atoms. They found the value 0.84087(39) fm, which is 4\% smaller than the value in the CODATA compilation at that time. Ever since, the muonic hydrogen atom and the anomaly in muon anomalous magnetic moment, the so-called $(g-2)$-factor, have stimulated the proposal of plenty of scenarios for seeking new physics beyond the Standard-Model~\cite{Lindner}. This is strongly justified because at the present days the difference between the experimental and the theoretical results is at the 3.3$\sigma$-level. Nowadays, muon and muonic atoms are a very rich laboratory for trying to understand new physics.

On the other hand, planar physics is a subject of increasing attention in connection with lower-dimensional Condensed Matter systems. Motivated by the present status of muon physics in connection with non-conventional Particle Physics scenarios and by the interest raised by special properties of planar systems in connection with new materials, we pursue, in this paper, an investigation of planar muonic atoms. Our starting point is the remarkable property that the planar electromagnetic interaction can be mediated by a massive photon of the Chern-Simons type, for which gauge invariance is not violated despite the photon becomes massive.

New materials like topological insulators, topological superconductors \cite{Liang,zhang,hsieh,chen,Xiao,Read,Xiao2,Liang2,Xiao3} and the genuinely planar graphene monolayers \cite{Geim,Geim2} bring with them the need to study what would be a new type of state of matter, which is generically called topological matter. To describe such states of matter, it is well known, at least theoretically, that Chern-Simons interaction plays an important role since, indeed, its formalism does provide effective field theories capable to describe them and other similar confined systems \cite{Kerler}.

In the framework of lower-dimensional Physics, more specifically planar systems, electromagnetic interaction allows that the photon acquires a non-trivial topological mass ($m_\gamma$) without any conflict with its inherent gauge symmetry. We are talking about the so-called Maxwell-Chern-Simons Electrodynamics \cite{CS-Deser1,CS-Deser2}, which, as quantum gauge-field theories, display a remarkable ultraviolet behavior and finds applications in the understanding and description of a great diversity of phenomena which are typical of (1+2) space-time dimensions (that is one temporal and two spatial dimensions), such as, for the example, the appearance of anyons \cite{CS-Leinaas,CS-Wilczek}, which obey an intermediate statistics and are present in many planar systems of interest \cite{Helayel, Helayel2, Helayel3}.

In this work, we are going to numerically solve six different two-dimensional atoms, $pe$, $de$, $te$, $p\mu$, $d\mu$, and $t\mu$ in two cases: first, when the system is submitted to Chern-Simon potential; the second, admitting that in a planar configuration the electromagnetic interaction is still given by the usual $1/r$ potential. Consequences of this Ansatz are discussed in Ref.~\cite{felipe2}. The alternative Ansatz, which corresponds to the choice of a logarithm potential, will not be considered here since it is well known that such a potential gives rise to a spectrum of positive atomic energies. The new results are compared with theirs three-dimensional analogs.

\section{Chern-Simons potential in brief}
For the classical Maxwell-Chern-Simons Electrodynamics in (1+2) dimensions, the magnetic and the electric fields of a pointlike charge $e$ are given (in the Gaussian unit system) by
\begin{equation}\label{campos}
  E=\frac{\hbar}{m_{\gamma} c}\nabla B \quad; \quad B=-V_0 \frac{m_{\gamma} c e}{2 \pi \hbar} K_0(m_{\gamma} c r/\hbar)
\end{equation}
where $K_0$ is the modified Bessel function of the second kind and $m_\gamma$ is the photon effective topological mass; $V_0$ is a factor that adjusts the energy dimension in two dimensions for the electrical potential.
The electrical potential associated to the electric field of Eq.~(\ref{campos}) is
\begin{equation}\label{chernp}
  \phi(r)=\frac{eV_0}{2\pi}K_0\left(\frac{m_{\gamma}c}{\hbar}r\right)
\end{equation}

Table~\ref{super} shows the expected values of $m_\gamma$ for different systems. In this paper we will use the following three values for $m_\gamma$: 1~eV, 10~eV and 100~eV.
\begin{table}[hbtp]
\caption{Range for possible values of $m_\gamma$}
\begin{center}
\begin{tabular}{|l|c|}
  \hline
  Superconductors  & $m_\gamma$  (eV)  \\ \hline \hline
  Conventional type I ({\tt Al, In, Sn, Pb, Nb})  & 0.1 -- 1  \\ \hline
  Conventional type I ({\tt Pb-In, Nb-Ti,l Nb-N, Pb-Bi})  & 2 -- 10  \\ \hline
  High temperature type II  & 10 -- 20  \\ \hline
  \hline
\end{tabular}
\end{center}
\label{super}
\end{table}

\newpage

\section{2D radial Sch\"{o}dinger equation}

The radial Schr\"{o}dinger equation in two dimensions for a generic  potential $V(r)$ is given by

\begin{equation} \label{eq: 9}
\frac{\mbox{d}^2u(r)}{\mbox{d}r^2}+\frac{2\mu}{\hbar^2}\Bigg[E-V(r)-\frac{\hbar^2}{2\mu}\frac{(\ell^2-\frac{1}{4})}{r^2}\Bigg]u(r)=0
\end{equation}
For the Coulomb potential, $\displaystyle V(r) = \frac{-e^2}{r}$, Eq.~(\ref{eq: 9}) can be rewritten in terms of the dimensionless variable $\rho$ as

\begin{equation} \label{1r2}
\frac{\mbox{d}^2u(\rho)}{\mbox{d}\rho^2}+\frac{a_B^2}{\zeta}\frac{2\zeta m_e}{\hbar^2}\Bigg[E+\frac{\sqrt{\zeta} e^2}{\rho a_B}-\frac{\hbar^2}{2\zeta m_e}\frac{\zeta(\ell^2-\frac{1}{4})}{\rho^2 a_B^2}\Bigg]u(\rho)=0
\end{equation}
with $\displaystyle r= \frac{\rho a_B}{\sqrt{\zeta}}$, $\mu = \zeta m_e$ and $a_B = \displaystyle \frac{\hbar^2}{m_e e^2}$ (where $m_e$ is the electron mass).

For Chern-Simons potential given by Eq.~(\ref{chernp}), \textit{i.e.}, $V(r) = -e\phi(r)$, Eq.~(\ref{eq: 9}) can be rewritten in terms of  $\rho$ as

\begin{equation}\label{csdim}
\frac{\mbox{d}^2u(\rho)}{\mbox{d}\rho^2}+\Bigg[\rho_0^2\frac{2 m_e}{\hbar^2} E+\frac{1}{\pi}K_0\left(\frac{\lambda m_{e}c}{\hbar}\frac{\rho \rho_0}{\sqrt{\zeta}}\right)-\frac{(\ell^2-\frac{1}{4})}{\rho^2}\Bigg]u(\rho)=0
\end{equation}

\noindent where $\displaystyle m_\gamma = \lambda m_e$, $\displaystyle \mu = \zeta m_e$, $\displaystyle r = \frac{\rho \rho_0}{\sqrt{\zeta}}$\ and \ $\displaystyle \rho_0 = \sqrt{\frac{\hbar^2}{m_e e^2 V_0}}$.

Let us consider, for simplicity, atomic units where $\hbar = m_e = e = 1$ and $c = \alpha^{-1} \simeq 137.0356$, with $\alpha$ being the usual fine structure constant. For $\rho_0$ to become our atomic unit of length, we must also consider $V_0 = 1$. Eqs.~(\ref{1r2}) and~(\ref{csdim}) can be written in the compact form
\begin{equation}\label{csfinal}
\fbox{$\displaystyle \frac{\mbox{d}^2u(\rho)}{\mbox{d}\rho^2}+\left[2E - U_{\text{eff}_i}\right]u(\rho)=0$}
\end{equation}

\noindent where we will use for the Coulomb potential the values
$$\displaystyle 2E = \varepsilon\ \, \text{and}\ \, \displaystyle \displaystyle U_{\text{eff}_1} = -\frac{2 \sqrt{\zeta}}{\rho}+\frac{(\ell^2-\frac{1}{4})}{\rho^2}$$
and, for the Chern-Simons potential,
$$\displaystyle 2E = \eta \ \, \text{and}\ \, U_{\text{eff}_2} = -\frac{1}{\pi}K_0\left(\frac{\lambda\rho}{\alpha\sqrt{\zeta}}\right) + \frac{(\ell^2-\frac{1}{4})}{\rho^2}$$

\begin{figure}[!ht]
\centerline{\includegraphics[width=9.cm]{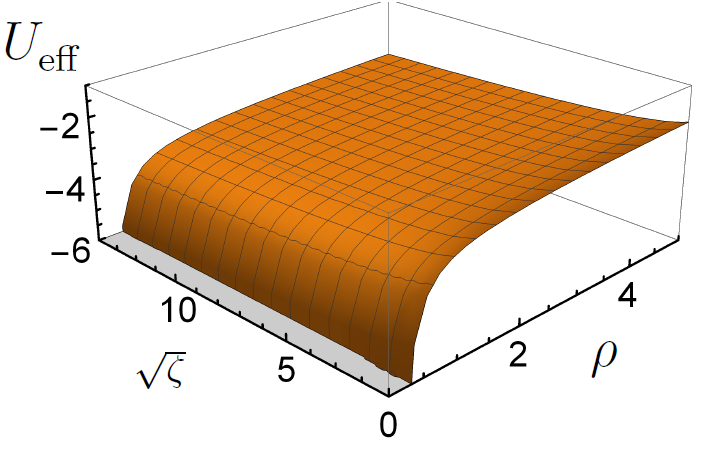}}
\caption{Effective potencial as a function of $\rho$ and $\sqrt{\zeta}$.}
\label{fig-eff}
\end{figure}

The effective potential $U_{\text{eff}_2}$ is plotted in Figure~\ref{fig-eff} as a function of $\rho$ and $\sqrt{\zeta}$, showing a smooth dependence on the parameter $\sqrt{\zeta}$.

Eq.~(\ref{csfinal}) can easily be adjusted for any neutral atom, just by changing the $\zeta$ value.
\newpage

\section{Numerical solutions}
To numerically solve the eigenvalue Eq.~(\ref{csfinal}), a slightly modified version of the Numerov method \cite{numerov, numerov2, numerov3, numerov4} will be used. The algorithm was implemented by the authors using C++ language. All the calculations were done with the CERN/ROOT package.

For the purpose of comparison we will also solve Eq.~(\ref{csfinal}) with the potential that describes these atoms in three-dimensional space, where $\displaystyle \beta = 2E$ and $\displaystyle U_{\text{eff}} = -\frac{2 \sqrt{\zeta}}{\rho} + \frac{ \ell(\ell+1)}{\rho^2}$.

With the particle masses of the atomic constituents given in Table~\ref{masses}, one can calculate the values of the parameter $\zeta$ ($\sqrt{\zeta}$) which are shown in Table~\ref{zeta}.

\renewcommand{\arraystretch}{1.2}
\begin{table}[ht]
  \caption{Mass values for different particles in terms of the electron mass}\label{tabelamassa}
  \vspace*{0.2cm}
  \begin{center}
  \begin{tabular}{|c|c|}
    \hline
      Particle    & Mass ($m_e$)     \\ \hline
      $e$         & 1              \\ \hline
      $\mu$       & 206.7682830    \\ \hline
      $p$         & 1836.15267343  \\ \hline
      $d$         & 3670.48296788  \\ \hline
      $t$         & 5496.92153573  \\
      \hline
  \end{tabular}
  \end{center}
  \label{masses}
\end{table}
\renewcommand{\arraystretch}{1}
\renewcommand{\arraystretch}{1.2}
\begin{table}[!ht]
  \caption{$\zeta$ values for different Atoms}\label{tabelazeta}
  \vspace*{0.2cm}
  \begin{center}
  \begin{tabular}{|c|c|c|}
    \hline
      Atom       & $\zeta$    & $\sqrt{\zeta}$   \\ \hline
      $pe$       & 0.99946    & 0.999730          \\ \hline
      $de$       & 0.99973    & 0.999865           \\ \hline
      $te$       & 0.99982    & 0.999910           \\ \hline
      $p\mu$     & 185.84083  & 13.632345         \\ \hline
      $d\mu$     & 195.74163  & 13.990769          \\ \hline
      $t\mu$     & 199.27259  & 14.116394           \\
      \hline
  \end{tabular}
  \end{center}
  \label{zeta}
\end{table}
\renewcommand{\arraystretch}{1}

Therefore, numerically solving Eq.~(\ref{csfinal}), we obtain for the ground state energies (in rydberg units, Ry) of the six different atoms the values shown in Table~\ref{tabelaenergia}. In the case of Chern-Simons potential the three different energies values correspond to different photon topological masses
which are, respectively, $\lambda = 0.2\times 10^{-5}$, $\lambda = 0.2\times 10^{-4}$ and $\lambda = 0.2\times 10^{-3}$.

Let us make here just two comments concerning the ground state energy of the hydrogen atom ($pe$) before showing some wave-functions. The first one is related to approximation of our numerical calculation. We can see from Table~\ref{tabelaenergia} that we found $\beta = -0.9833$~Ry instead of the well known result $\beta = -1$~Ry, meaning that they are compatible with a numerical error of about 1.67\%. The second point regards a previous prediction~\cite{jordan}. There, for a planar system, the ground state energy $\eta$ for the Chern-Simons interaction with different $m_\gamma$ values was found to be in the following range: $-3.5 \times 10^{-3}~\text{eV} \lsim \eta \lsim - 9.0 \times 10^{-2}~\text{eV}$ or, equivalently, $-4.67 \times 10^{-2}~\text{Ry} \lsim \eta \lsim 1.22~\text{Ry}$. So, the conclusion of that paper was that in 2D the hydrogen atom is weekly bounded.
\renewcommand{\arraystretch}{1.2}
\begin{table}[hbt]
  \caption{Ground state ($\ell=0$) energy for different atoms considering different potentials. Here $\beta$ is the ground state energy for the 3D atom, $\varepsilon$ is the ground state energy for the 2D atom with the usual $1/r$ Coulombic potential and $\eta$ is the ground state energy for the 2D atom with the Chern-Simons Potential.}
  \begin{center}
  \begin{tabular}{|c|c|c|c|c|c|}
    \hline
      Atom        & $\beta$       & $\varepsilon$    &  \multicolumn{3}{c|}{$\eta$}        \\ \hline
      $pe$        &   -0.9833     &   -2.1           &  -2.2417     &  -1.5083     &  -0.775          \\ \hline
      $de$        &   -0.9833     &   -2.1           &  -2.2417     &  -1.5083     &  -0.775          \\ \hline
      $te$        &   -0.9833     &   -2.1           &  -2.2417     &  -1.5083     &  -0.775          \\ \hline
      $p\mu$      &   -185.8      &   -401.88        &  -3.0667     &  -2.3333     &  -1.6            \\ \hline
      $d\mu$      &   -194.67     &   -411.47        &  -3.0750     &  -2.3417     &  -1.6083         \\ \hline
      $t\mu$      &   -199.13     &   -410.28        &  -3.0833     &  -2.35       &  -1.6167         \\
      \hline
  \end{tabular}
  \end{center}
  \label{tabelaenergia}
\end{table}
\renewcommand{\arraystretch}{1}

A direct  inspection of Table~\ref{tabelaenergia} shows that our prediction is in the interval $-2.2417~\text{Ry} \lsim \eta \lsim -0.775~\text{Ry}$. This actually means that a planar hydrogen atom with Chern-Simons interaction is still more bounded than that in 3D, except for the limit value $\lambda = 0.2\times 10^{-3}$. However, even in this case, the results are quite close.

The discrepancy between our result and that of Ref.~\cite{jordan} is due to a difference in the effective potential. Indeed the potential used in that paper is not correct. In our notation it is given by
$$U_{\text{eff}} = -\frac{\lambda}{\pi\alpha}K_0\left(\frac{\lambda\rho}{\alpha\sqrt{\zeta}}\right) + \frac{(\ell^2-\frac{1}{4})}{\rho^2}$$
to be compared to our potential
$$U_{\text{eff}} = -\frac{1}{\pi}K_0\left(\frac{\lambda\rho}{\alpha\sqrt{\zeta}}\right) + \frac{(\ell^2-\frac{1}{4})}{\rho^2}$$
The multiplicative factor in front of $K_0$ forces the results to be quite smaller than those found in the present paper.

\begin{figure}[!ht]
\centerline{\includegraphics[width=7.0cm]{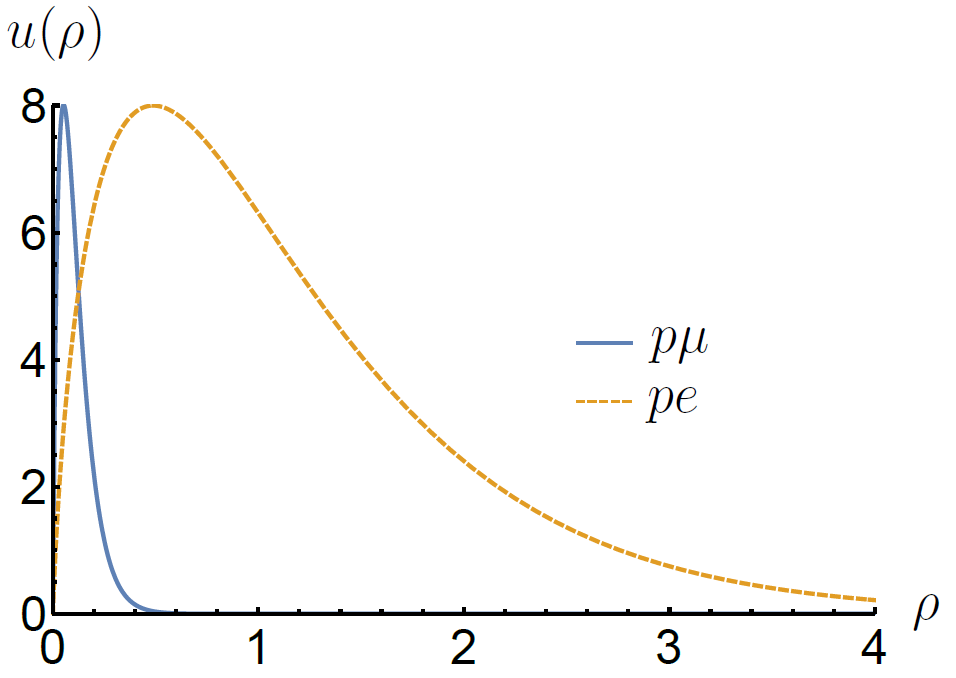}}
\caption{Ground state wave function for two different atoms with the coulombic potential $-e^2/r$; $p\mu$ is represented by the continuous line and $pe$ is represented by the dashed line.}
\label{fig-a}
\end{figure}

\begin{figure}[!ht]
\centerline{\includegraphics[width=7.0cm]{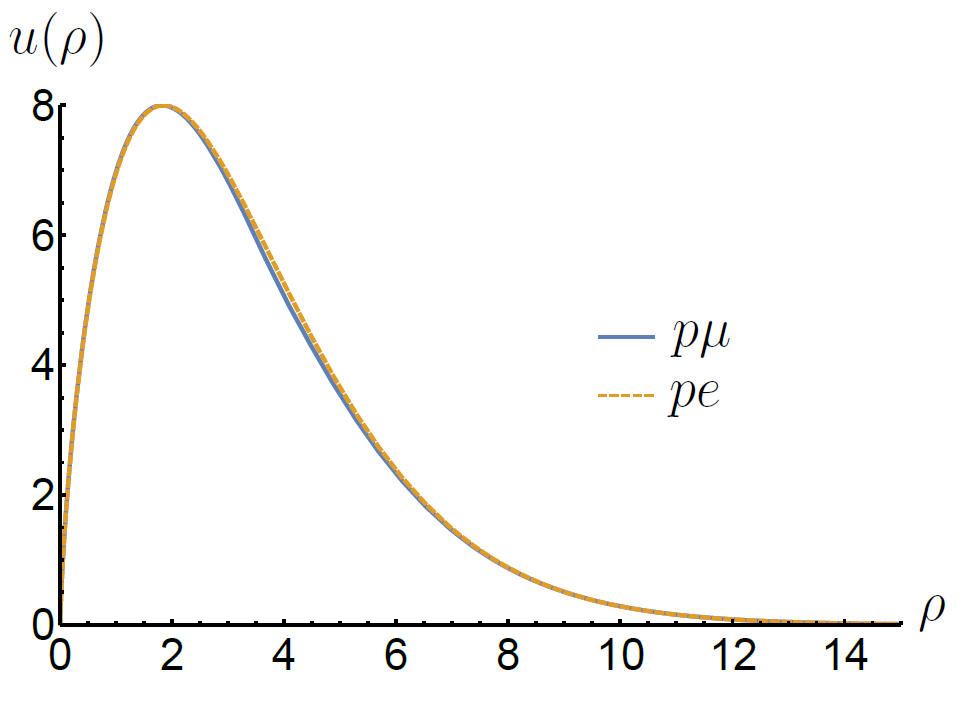}}
\caption{Ground state wave function for two different atoms with the Chern-Simons potential, $p\mu$ is represented by the continuous line and $pe$ is represented by the dashed line.}
\label{fig-b}
\end{figure}

Let us now reproduce some wave-functions.
For simplicity, we will show the comparison of the wave-functions only for some of the calculated atoms. Fig.~\ref{fig-a}, for example, shows the ground state wave functions for the Eq.~(\ref{1r2}) for two different atoms ($pe$ and $p\mu$). In Fig.~\ref{fig-b} it is shown the $l=0$ solutions of Eq.~(\ref{csfinal}) for the same atoms.

\newpage
We also calculate numerically the mean atomic radius $\langle r \rangle$ in terms of the Bohr radius for some of the studied atoms, which is given by

\begin{equation*}
 \langle r \rangle = \frac{\int r^D R^2_{\ell=0}(r)\mbox{d}r}{\int r^{D-1} R^2_{\ell=0}(r)\mbox{d}r}
\end{equation*}
\noindent with $D$ representing the dimension to be used in the calculation and $R_\ell(r)$ is the radial wave-function which is solution of Eq.~\ref{csfinal}. From the above equation, we got the values reported in Table~\ref{mean-radius}.

\renewcommand{\arraystretch}{1.0}
\begin{table}[hbt]
  \caption{Mean radius (in unit of Bohr's radius) of some atoms for the 3D $1/r$ potential, the 2D $1/r$ potential and the Chern-Simons potential with $\lambda = 0.2\times 10^{-4}$}
  \begin{center}
  \begin{tabular}{|c|c|c|c|}
    \hline
      \multirow{2}{*}{Atom}      &   \multicolumn{3}{c|}{$\langle r \rangle/a_B$}         \\
                                 &   \multicolumn{3}{c|}{3D \qquad    2D  \qquad  \quad  CS}      \\ \hline
            $pe$                 &1.5        &0.8331       &2.58107   \\ \hline
            $p\mu$               &0.00807    &0.0044       &0.188879   \\ \hline
            $t\mu$               &0.00753    &0.0042       &0.184474   \\
      \hline
  \end{tabular}
  \end{center}
  \label{mean-radius}
\end{table}
\renewcommand{\arraystretch}{1}

Notice that the mean atomic radius for the Chern-Simons interaction is always the largest one.

We have computed the eigenvalues of Eq.~\ref{csfinal}, in the cases $\ell=1$ and $\ell=2$, for all the six atoms listed in Table~\ref{tabelaenergia}, corresponding to $\ell=0$. However, in Table~\ref{l1l2} we reproduce the results for $\ell=1,2$ just for $pe$ and $p\mu$ atoms.

\renewcommand{\arraystretch}{1.0}
\begin{table}[hbt]
  \caption{Ground state energy for $\ell=1$ and $\ell=2$ for two different atoms ($pe$ and $p\mu$) with the Chern-Simons potential and $\lambda = 0.2\times 10^{-5}$}
  \begin{center}
  \begin{tabular}{|c|c|c|}
    \hline
             Atom                &$\ell=1$   &$\ell=2$  \\ \hline
            $pe$                 &-2.017     &-1.95     \\ \hline
            $p\mu$               &-2.85      &-2.783     \\
      \hline
  \end{tabular}
  \end{center}
  \label{l1l2}
\end{table}
\renewcommand{\arraystretch}{1}

\section{Discussion}

\begin{figure}[!ht]
\centerline{\includegraphics[width=8.0cm]{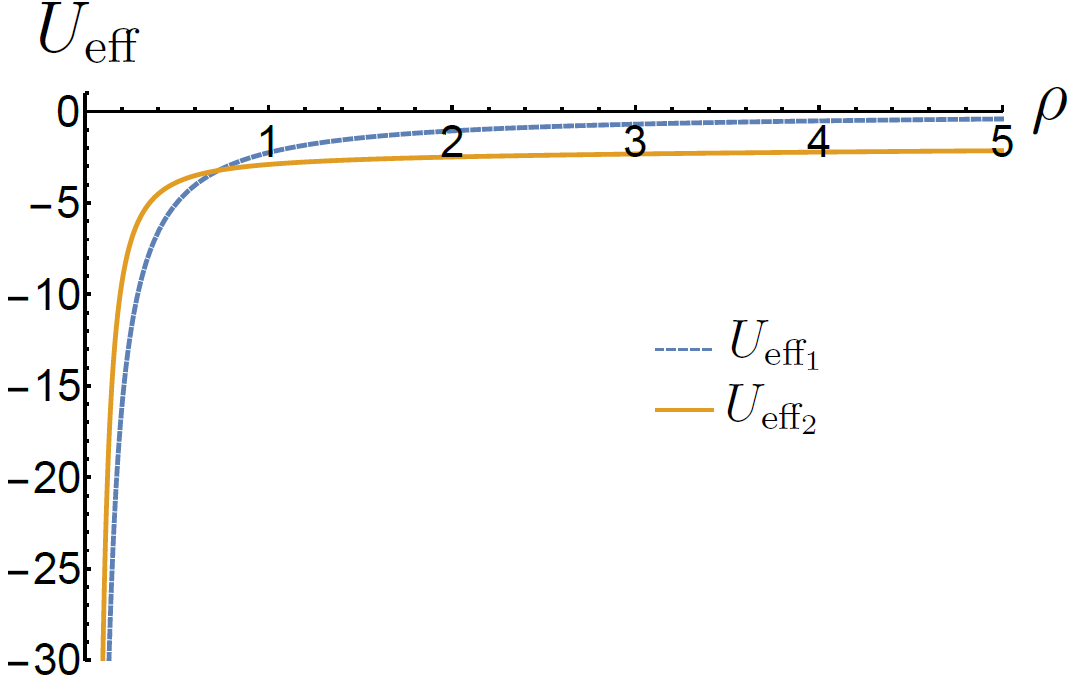}}
\caption{Dashed line represents the effective potential shown in Eq.~(\ref{1r2}) and the continuous line is the effective potential shown in Eq.~(\ref{csfinal}), both of them correspond to the atom $pe$, for $\ell=0$.}
\label{fig-pe}
\end{figure}

\begin{figure}[!ht]
\centerline{\includegraphics[width=8.0cm]{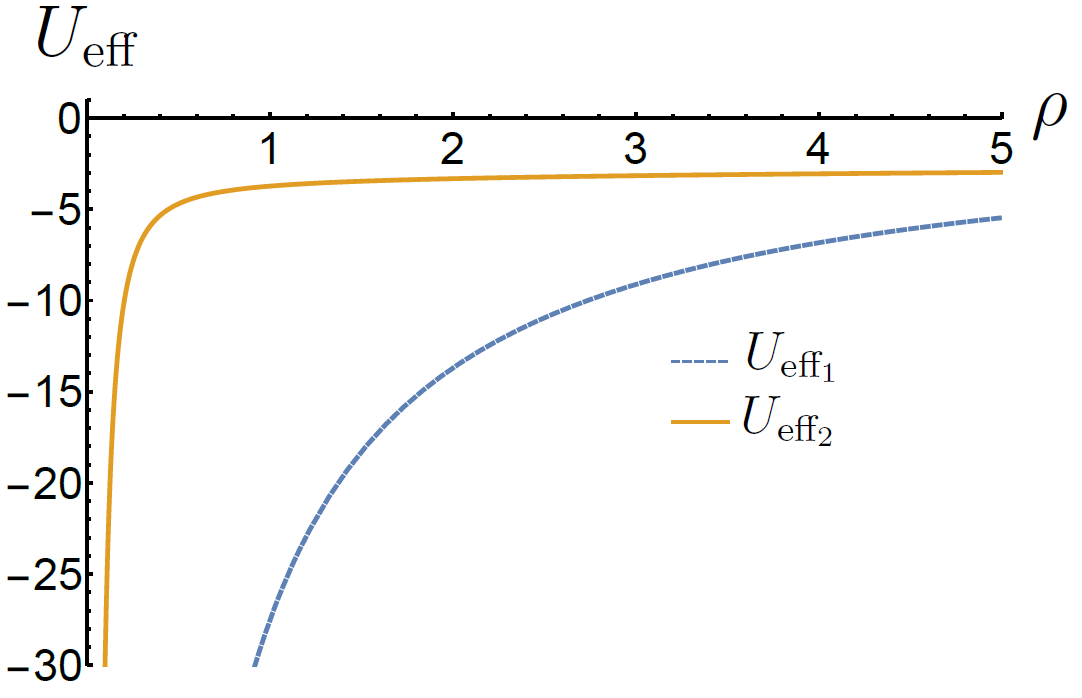}}
\caption{Dashed line represents the effective potential shown in Eq.~(\ref{1r2}) and the continuous line is the effective potential shown in Eq.~(\ref{csfinal}), both of them correspond to the atom $p\mu$, for $\ell=0$.}
\label{fig-pmu}
\end{figure}

For a better understanding of the results presented in the previous Section, we will graphically compare the two effective potentials used in the Eqs.~(\ref{1r2}) and~(\ref{csfinal}) for $pe$ atom (Fig.~\ref{fig-pe}) and for $p\mu$ atom (Fig.~\ref{fig-pmu}).

First of all, we can see that the Chern-Simons effective potential doesn't change much when we move from an electron to a muon just by comparing Figs.~\ref{fig-pe} and \ref{fig-pmu}. This explains why in Table~\ref{tabelaenergia} the energy of the electronic atoms is so close to the energy of muonic atoms. Thus, even though the muon is about 200 times more massive than the electron, the dependence of the result on topological mass of the photon still dominates the equation.

Thus, let us see how the potential varies according to the choice of topological mass value of the photon in the $pe$ atom interaction.

\begin{figure}[!ht]
\centerline{\includegraphics[width=8.0cm]{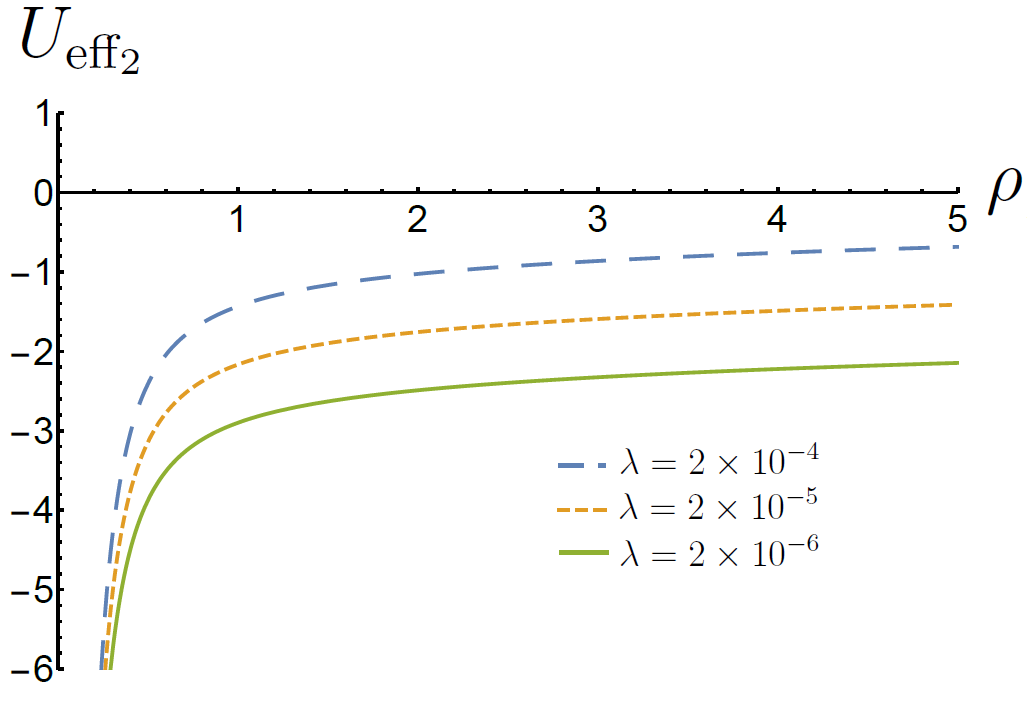}}
\caption{Chern-Simons effective potential for 3 different values of the scale parameter $\lambda = m_{\gamma}/m_e$.}
\label{fig-c}
\end{figure}
\begin{figure}[!ht]
\centerline{\includegraphics[width=9.0cm]{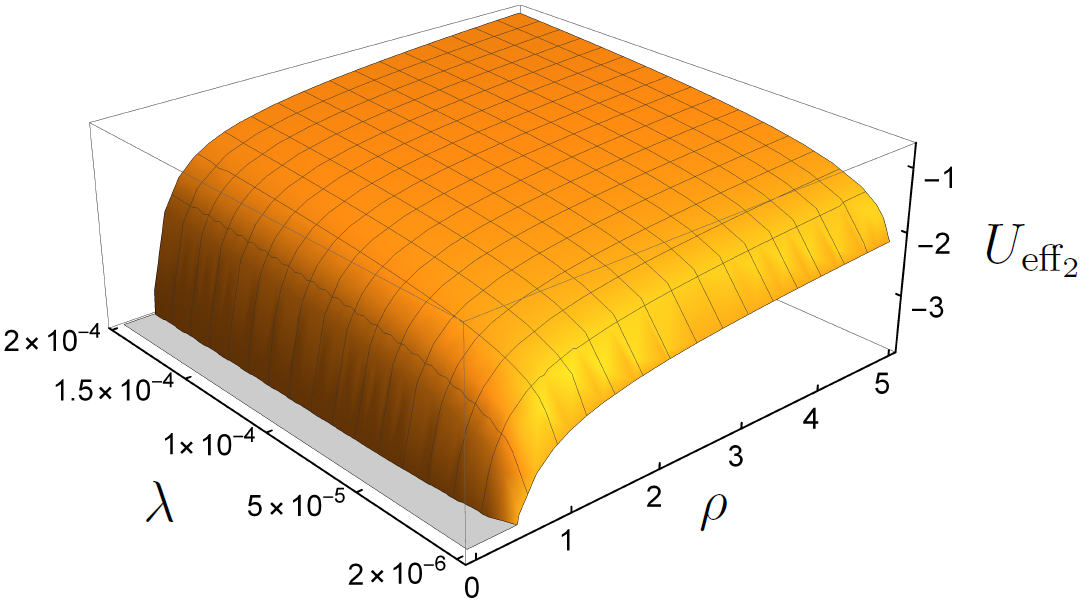}}
\caption{Chern-Simons effective potential for different values of $\lambda$.}
\label{fig-d}
\end{figure}

Analyzing Fig.~\ref{fig-c} and \ref{fig-d} together with the data in Table~\ref{tabelaenergia}, we can see that the greater the topological mass of the photon, the greater the energy of the atom's ground state.

Comparing our predictions with those reported in another paper~\cite{jordan}, we observe, as already stressed, a significant discrepancy in the ground state energy that comes to be of the order of $10^4$ and of the order of 10 to $10^2$ for the atomic mean radius. This fact is due to a mistake in the two-dimensional electrical Chern-Simons potential. Therefore, our results demonstrate that it is not true that Chern-Simons interaction gives rise to atomic configurations that are similar to Rydberg atoms in 3D.

\newpage

\section{Conclusion}
We have shown that an electric neutral system governed by the Chern-Simons potential does not present an important characteristic usually attributed to muonic atoms, namely the fact that, being 200 times heavier than the electron, the muon inside an atom reduces significantly the distance between nucleons, bringing the atom to a more bounded state than its analog with an electron. This fact is not a good news for certain applications, such as the muon catalyzed fusion \cite{felipe3}, which is based on the general idea that the muon atom has a much smaller internuclear separation than the electronic one. Indeed, from Table~\ref{mean-radius}, we see that, although Chern-Simons interaction gives rise to smaller radius, the mean radius for both $p\mu$ and $T\mu$ are still about 25 times greater than the equivalent atom governed by a Coulomb potential in 3D.

Finally, a comparison the results obtained in this paper with experimental data may shed light on the real interaction that occurs in almost two-dimensional systems. Once one can perform experiments on planar materials with muonic and electronic atoms, their experimental differences can be used to rule out, or not, the presence of a Chern-Simons interaction.

\section*{Acknowledgment}

One of us (FS) was financed in part by the Coordena\c{c}\~{a}o de Aperfei\c{c}oamento de Pessoal de N\'{\i}vel Superior -- Brazil (CAPES), Finance Code 001.

\bibliography{mybibfile}
\end{document}